\documentclass{aa}  

\usepackage{graphicx}
\usepackage{placeins}
\usepackage{float}
\usepackage{txfonts}
\usepackage{hyperref}
\hypersetup{
    colorlinks=true,
    linkcolor=blue,
    filecolor=magenta,      
    urlcolor=cyan,
    citecolor=blue
    }

\def\lcdm{$\Lambda$CDM}
\def\S8{$S_8$}
\def\kg{$\kappa_g$}
\def\omegm{$\Omega_m$}
\def\sig8{$\sigma_8$}

\begin{document}

   \title{Interpretability of deep-learning methods applied to large-scale structure surveys}
   \author{G. Aymerich\inst{1,2} \fnmsep\thanks{\email{gaspard.aymerich@universite-paris-saclay.fr}} \and
   T. Kacprzak\inst{3,4} \and
   A. Refregier\inst{3}
          }

          \institute{
            Université Paris-Saclay, CNRS, Institut d'Astrophysique Spatiale, 91405, Orsay, France
   \and
            Université Paris-Saclay, Université Paris Cité, CEA, CNRS, AIM, 91191, Gif-sur-Yvette, France   
   \and
            Institute for Particle Physics and Astrophysics, ETH Zurich, 8093 Zurich, Switzerland
   \and
            Swiss Data Science Center, Paul Scherrer Institute, 5232 Villigen, Switzerland}

   \date{Received XXX; accepted YYY}

 
  \abstract{
  Deep learning and convolutional neural networks in particular are powerful and promising tools for cosmological analysis of large-scale structure surveys. They are already providing similar performance to classical analysis methods using fixed summary statistics, are showing potential to break key degeneracies by better probe combination and will likely improve rapidly in the coming years as progress is made in the physical modelling through both software and hardware improvement. One key issue remains: unlike classical analysis, a convolutional neural network's decision process is hidden from the user as the network optimises millions of parameters with no direct physical meaning. This prevents a clear understanding of the potential limitations and biases of the analysis, making it hard to rely on as a main analysis method. In this work, we explore the behaviour of such a convolutional neural network through a novel method. Instead of trying to analyse a network "a posteriori", i.e. after training has been completed, we study the impact on the constraining power of training the network and predicting parameters with degraded data where we removed part of the information. This allows us to gain an understanding of which parts and features of a large-scale structure survey are most important in the network's prediction process. We find that the network's prediction process relies on a mix of both Gaussian and non-Gaussian information, and seems to put an emphasis on structures whose scales are at the limit between linear and non-linear regimes.
  }


   \maketitle
%

\section{Introduction}
\label{intro}
The Lambda-Cold Dark Matter (\lcdm) model has been the standard model of cosmology for the past 20 years, and has been able to explain a wide range of observations, from the cosmic microwave background anisotropies to the large-scale structure of the Universe. However, the model is not without its issues: not only are the nature of dark matter and dark energy, the two components that make up 95\% of the Universe, still unknown, but the model has also been unable to coherently explain some observations, such as the tension between the Hubble constant measurements obtained by the Planck collaboration and the SH0ES collaboration \citep{planck_collaboration_planck_2020,riess_comprehensive_2022}, or the tension between the $S_8 \equiv \sigma_8 \sqrt{\Omega_m / 0.3}$ values found by the Planck collaboration and most large-scale structure surveys \citep{planck_collaboration_planck_2020, heymans_kids-1000_2021, abbott_dark_2022}.
In order to better understand those limitations of the standard cosmological model, next generation surveys such as the Euclid space telescope, the Vera Rubin Observatory, or the Simons Observatory are under construction or already collecting data. The quality, complexity, and volume of the data sets expected from those surveys is such that the statistical tools used for data analysis may become the limiting factor of future studies. Indeed, it is impossible to directly compare the raw data from a survey to the theoretical models of the Universe, and one has to compress the information in some way to obtain constraints on the model parameters. This step is crucial and the data compression has to be chosen carefully to avoid information loss, which is not straightforward in most cases. In this work, we focus on the analysis of large-scale structure surveys that measure the distortion of the shape of distant galaxies by the presence of mass between these galaxies and the observer in order to directly probe the distribution of matter (both ordinary and dark matter) in the Universe \citep[see][for a review on weak lensing]{refregier_weak_2003}.\\

\subsection{Summary statistics}
\label{summary_stats}

Weak lensing surveys such as the Dark Energy Survey (DES) or the Kilo-Degree Survey (KiDS) measured the shape and position of hundreds of millions of galaxies to map the lensing shear over large areas of the sky. To obtain constraints on the \lcdm\ model parameters, these maps need to be compared to the model's predictions which can be done via the computation of summary statistics such as the power spectrum, higher-order moments, peak count, or Minkowski functionals. An intrinsic limitation of this summary statistics approach stems from the fact that the shear field probes the late-time Universe where non-linearities have arose. Therefore, unlike the cosmic microwave background anisotropies that can be very well modelled by a Gaussian random field for which the power spectrum can encapsulate the entire information \citep[see e.g.][]{planck_collaboration_planck_2014-2, aiola_atacama_2020, balkenhol_measurement_2023}, no single summary statistics can capture the entire information contained in the shear maps, and the "classical" approach of analysis via summary statistics is bound to result in some information loss \citep{bernardeau_weak_1997,villaescusa-navarro_neural_2022}. Extensive work has been done on the theoretical modelling of the Universe beyond the 2-point correlation function or power spectrum in order to access the non-Gaussian information of cosmological surveys \citep[see e.g.][]{takada_three-point_2003,takada_three-point_2003-1,friedrich_density_2018,gong_cosmology_2023,heydenreich_roadmap_2023}, but the actual data analysis remains challenging, and most baseline analyses of weak-lensing surveys only leverage 2-point correlation functions. Nevertheless, using these methods, the KiDS and DES collaborations managed to measure cosmological parameters with under 5\% precision \citep{heymans_kids-1000_2021, abbott_dark_2022}. Given the fact that lensing surveys are affected by astrophysical effects such as galaxy intrinsic alignment and galaxy bias \citep{kirk_cosmological_2012, amon_dark_2022}, and the existence of a tension between the $S_8 \equiv \sigma_8 \sqrt{\Omega_m / 0.3}$ measurements obtained by late-time (such as lensing) and early-time (such as primary CMB anisotropies) probes, extracting the full information from the large-scale structure surveys is crucial. This problem will be exacerbated by the next generation surveys such as the Vera Rubin Observatory or the Euclid space telescope, which will provide much larger data sets, probing even smaller scales where non-linearities dominate, requiring the extraction of non-Gaussian information.\\

\subsection{Deep learning}
\label{deeplearning}
In recent years, machine learning methods are becoming increasingly popular in the field of cosmology and astrophysics, and have been adapted to a variety of different tasks, from classification to acceleration of simulations via emulators \citep[see][for a review of machine learning techniques in cosmology]{dvorkin_machine_2022}. In particular, deep-learning techniques have opened up a new path for obtaining constraints on the cosmological parameters from surveys. This new approach, which is a form of simulation-based inference, has been developed to analyse surveys (including, but not limited to, large-scale structure surveys) without the information loss inherent to fixed summary statistics methods: through the use of convolutional neural networks (CNN), which are a type of neural networks particularly suited to image analysis and pattern recognition \citep{lecun_backpropagation_1989}, one can directly input the raw survey data into the network and obtain cosmological parameters estimation as an output. In the case of weak-lensing surveys, the CNN can be trained using simulated lensing and galaxy density maps obtained with N-body simulations of the Universe for various cosmologies. This approach is still at an early stage and will likely benefit from the rapid improvement of both physical modelling as well as deep-learning-based image analysis in the future, but results are already showing the potential of this method and are consistent with the previous results obtained by other methods on the same data sets \citep{ravanbakhsh_estimating_2016,fluri_cosmological_2019,fluri_full_2022,pan_cosmological_2020-1, lu_cosmological_2023}. Recent work has shown that a deep-learning-based approach could help break degeneracies that currently limit our ability to account for astrophysical uncertainties, especially the galaxy intrinsic alignment/\S8 degeneracy, by allowing for a better combination of probes (shear and galaxy density in this case) than a 2-point cross-correlation function \cite{kacprzak_deeplss_2022}.\\
While this simulation-based inference approach shows a lot of promise for future work, there is one considerable drawback: unlike more traditional methods, the neural networks used for the data analysis are black boxes and it is hard to understand where the information is coming from. In the machine-learning community, this issue is know as the interpretability of machine learning models and is an active research field. Interpretability in the broad domain of machine learning in general is an active field of research, with many studies proposing methods to understand the behaviour of neural networks \citep[see][for a review]{samek_explaining_2021}. In the specific context of cosmology, interpretability is a very recent area of research, but a few studies have tried to tackle this problem with different approaches \citep[see e.g.][]{matilla_interpreting_2020,villanueva-domingo_removing_2021,piras_representation_2024,lucie-smith_deep_2024,gong_c3nn_2024}. In this work, we will use a simplified version of the DeepLSS architecture presented in \cite{kacprzak_deeplss_2022} and try to understand where the information extracted by the neural network is coming from.\\
\\
The paper is organized as follows: in Sect.\ref{data}, we present the data set and network architecture used in this work. In Sect.\ref{interpretability}, we present the approach used to understand the network's behaviour and decision process. In Sect.\ref{results}, we present the results obtained with this framework. Finally, in Sect.\ref{conclusion}, we summarise and compare our results with other studies.

\section{Data set and network architecture}
\label{data}
We use the same approach as in \cite{kacprzak_deeplss_2022} with a simpler setup to speed up the training process as this work includes a large number of models to train, and is not focused on testing the maximum performance potential of the network architecture but rather understanding how it works. As both the data set and network architecture are adapted from previous work, this section will be brief and focused on explaining how they were simplified. We refer the interested reader to \cite{kacprzak_deeplss_2022} for more details.

\subsection{Weak lensing maps}
\label{maps}
\begin{figure}
\includegraphics[width=\columnwidth]{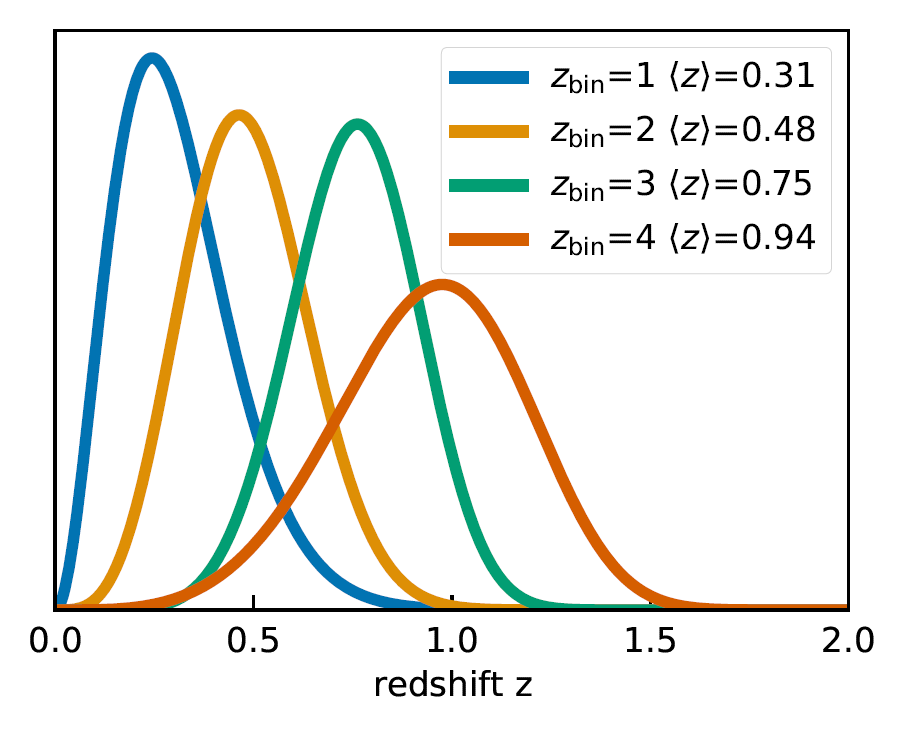}
\caption{Redshift bins used for this work, chosen to be generally representative of a Stage III survey. Figure taken from \cite{kacprzak_deeplss_2022}.}
\label{fig:z_bin}
\end{figure}

We use a set of tomographic weak lensing convergence \kg\ maps (this work doesn't include galaxy clustering data) created for a flat \lcdm\ model with a fixed dark energy equation of state $w=-1$. We vary the value of the cosmological parameters \omegm\ and \sig8\ within $[0.15,0.45]$ and $[0.5,1.2]$ respectively and set the other parameters to the baseline results of \cite{planck_collaboration_planck_2020}: $\Omega_b=0.0493$, $H_0=67.36$, $n_s=0.9649$. As the goal of this work is understanding the network's behaviour and not matching the physical reality as closely as possible, measurement uncertainties like redshift errors or selection function are not taken into account. We also do not include galaxy intrinsic alignment in the baseline analysis, but the impact of its inclusion is studied in Appendix \ref{intrinsic}.\\
We use a survey configuration with 900 deg² and 2.5 galaxies/arcmin² distributed evenly across 4 redshift bins with mean redshift $\langle z \rangle =0.31,0.48,0.75,0.94$ (see Fig.\ref{fig:z_bin}).
We follow the method introduced in \cite{fluri_cosmological_2019} to calculate convergence \kg\ maps. Using the \texttt{PKDGRAV3} code \citep{potter_pkdgrav3_2017}, 57 unique cosmologies in the \omegm -\sig8 plane were each simulated 12 times. Each simulation used $256^3$ particles in a volume of $500^3$ Mpc$^3$, and the initial conditions were generated at redshift $z_{init} = 50$, using the \texttt{MUSIC code} \citep{hahn_multi-scale_2011}. All simulations were run over 500 time steps, saving snapshots at intervals of $\Delta z = 0.1$ from $z = 3.45$ to $z = 1.55$ and $\Delta z = 0.05$ from $z = 1.55$ to $z = 0$. From the 3D simulated overdensity $\delta_{3D}$, 2D convergence maps are projected using the 
\texttt{UFALCON} code \citep{sgier_fast_2019}. The convergence of a given pixel is calculated with the Born approximation:
\begin{figure*}
\includegraphics[width=\textwidth]{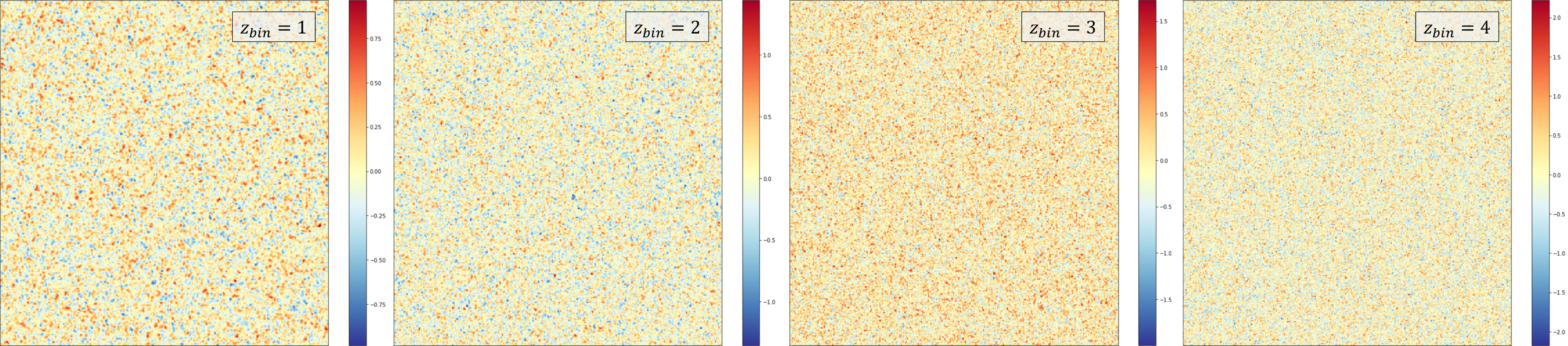}
\caption{Example of a simulated 900 deg² survey with 4 redshift bins, obtained by creating a 
mosaic of $6\times6$ individual $5\times5$ degrees maps, with Gaussian noise and Gaussian smoothing at scale $R=4$ Mpc/h}
\label{fig:maps}
\end{figure*}
\begin{equation}
   \label{eq:born_approx}
\kappa \approx \sum_{b}^{}W^{WL}_{b} \int_{\Delta z_b}^{} \frac{dz}{E(z)} \, \delta_{3D} [ \frac{c}{H_0} \mathcal{D}(z) \hat{n}^{pix},z]   
\end{equation}
where $\mathcal{D}(z)$ is the dimensionless comoving distance, $E(z)$ is defined by $d\mathcal{D}=\frac{dz}{E(z)}$ and $\hat{n}^{pix}$ is a unit vector pointing to the pixel. The sums runs over the redshift shells (that are of thickness $\Delta z_b$) and the weight for each shell is defined as:
\begin{equation}
   \label{eq:weights}
W^{WL}_{b} = \frac{3}{2} \Omega_m \frac{\int_{\Delta z_b}^{} \frac{dz}{E(z)} \int_{z}^{z_s} dz' n(z') \frac{\mathcal{D}(z)\mathcal{D}(z,z')}{\mathcal{D}(z')a(z)}} {\int_{\Delta z_b}^{} \frac{dz}{E(z)} \int_{z_0}^{z_s} dz' n(z')}  
\end{equation}
The mean convergence of each map is subtracted:
\begin{equation}
   \label{eq:mean_removal}
   \kappa \leftarrow \kappa - \langle\kappa \rangle
\end{equation}
The maps are noised over with Gaussian noise:
\begin{equation}
   \label{eq:noise}
\kappa_g = \textrm{Normal}\left[ \kappa, \frac{\sigma_e}{\sqrt{n_{gal}}}\right]
\end{equation}

where $\sigma_e=0.4$ is the galaxy shape noise and $n_{gal}$ is the number of galaxy per pixel.\\
Because the simulation are dark matter only, the physical modelling does not accurately describe the small scales where baryonic effects become apparent. Like in \cite{porredon_dark_2022} we use a smoothing scale of $R=4$ Mpc/h to discard the smallest scales. Taking the pixel size into account, we apply the following additional Gaussian smoothing for the four redshift bins:
\begin{equation}
   \label{eq:smoothing}
\sigma=[4.8,\ 3.5,\ 2.8,\ 2.5] \textrm{arcmin}
\end{equation}
Each individual map obtained through this process is $5\times5$ degrees and $64\times64$ pixels. We construct simulated 900 deg² surveys by creating a mosaic of $6\times6$ maps. Noise is added on-the-fly for each realization during both training and prediction, and maps are randomly placed and flipped when creating the mosaic to avoid repetition. Smoothing is done before the mosaic is created to avoid blurring the edges between maps. Fig.\ref{fig:maps} shows an example of a simulated survey.

\subsection{Neural networks}
\label{networks}
Most of the map analysis in this work is done using a convolutional neural network (CNN) that takes pixel maps as inputs and returns summary statistics corresponding to cosmological parameters \omegm\ and \sig8.The CNN is based on a ResNet architecture \citep{he_deep_2015} and consists of 4 convolutional layers, 10 residual layers with a kernel size of 5, a stride of 2 and the Relu activation function. The last residual layer is flattened and fully connected to the output layer that contains the summary statistics and their covariances. This architecture gives a total of $\sim10^7$ trainable parameters.\\
A power spectrum (PS) approach is also used, by calculating the auto- and cross-spectra of the input maps and compressing the PS vectors into the same summary statistics as the CNN via a neural network. Conducting power spectrum analysis this way was chosen to allow for simultaneous training of both networks and for the same likelihood analysis to be used on both CNN and PS outputs.\\
The power spectra of the \kg\ maps are calculated by FFT and averaged in 20 logarithmically spaced $\ell$-bins over the range $\ell \in [36,4536]$. This gives a minimum interval of $\delta \ell = 36$ which is the resolution of the FFT. Calculating all auto- and cross-spectra over the four redshift bins gives 10 spectra per simulated survey, which are then given as input to a network consisting of 2 fully connected layers with 1024 units and Relu activation, and an output layer identical to that of the CNN predicting summary statistics and covariances. This gives $\sim1.2 \, 10^6$ trainable parameters. The training strategy for both network architectures uses a negative log-likelihood loss function for both network types:
\begin{equation}
   \label{eq:loss}
L = \ln(|\Sigma_p|) + (\theta_p - \theta_t)^\top \Sigma_p^{-1} (\theta_p - \theta_t)
\end{equation}
where $\theta_t$ is the true parameter vector, $\theta_p$ is the output summary vector and $\Sigma_p$ is the output covariance matrix. The networks are trained using stochastic gradient descent with the \texttt{ADAM} optimizer \citep{kingma_adam_2017}, with a batch size of 32 simulated surveys and a learning rate of 0.00005 for the CNN and 0.0025 for the PS neural network. We additionally apply gradient clipping using the method described in \cite{seetharaman_autoclip_2020}, using 50\% percentile. Both networks (including hyper-parameters) are directly taken from \cite{kacprzak_deeplss_2022}, are created in \texttt{TensorFlow} \citep{abadi_tensorflow_2016} and trained on the CSCS supercomputer Piz Daint, using NVIDIA Tesla P100 16GB GPUs.\\
All models are trained for 100k batches, with very minimal improvement of the loss for the last 30k batches which indicates reasonable convergence. We create a separate test set consisting of around 8\% of the full data set that wasn't used for training and calculated the loss for this set throughout training. No difference between training and testing loss appears for any model, indicating that over-fitting was not a problem, most likely due to the addition of noise on each realization.

\subsection{Likelihood analysis}
\label{likelihood}
The conditional probability distribution $p(\theta_p|\theta_t)$ is estimated by running a prediction through the trained networks with samples for all 57 combinations of (\omegm,\sig8) from the simulation grid. $p(\theta_p|\theta_t)$ is then modelled via a Mixture Density Network (MDN) that uses a mixture of Gaussians at each $\theta_t$ to mimic the conditional probability distribution as closely as possible. The MDN predicts the means, covariances and relative weights of its Gaussian components, and is trained with stochastic gradient descent. The validation loss is monitored to prevent over-fitting and we confirm that the MDN modelled the probability distribution correctly with 4 Gaussian components by directly comparing the two.\\
We then run Markov Chain Monte Carlo (MCMC) with the \texttt{emcee} algorithm \citep{foreman-mackey_emcee_2013} with the distribution given by the MDN. 200 chains of 128k samples are run for each model (or a single 1.28m chain for plotting Fig.\ref{fig:triangleplot}).

\section{A novel approach to the interpretability problem}
\label{interpretability}
The unsimplified version of this network architecture presented in \cite{kacprzak_deeplss_2022} shows the great potential that deep-learning has to improve the analysis of weak-lensing surveys in the future, but it is hard to trust a "black-box" analysis method whose limitations and potential biases are unknown and we also cannot optimize our future surveys for it should it become the main analysis method. It is therefore crucial to try to understand how the CNN works and where the information is coming from.\\
As more and more effort is put into developing deep-learning approaches to weak lensing data analysis and great potential is starting to appear, ways to solve the interpretability problem are starting to be pursued. This work chooses a novel approach (at least in the field of weak-lensing): instead of using "a posteriori" interpretability methods like many studies where the network is first trained on the normal data set then analysed through various methods, such as sensitivity maps which measure how much the output is affected by a variation of each input pixel \citep{lecun_gradient-based_1998}, we chose to directly degrade the data set then train the networks with the modified data and see how much the constraining power is affected compared to a network trained with the full data.
\begin{figure*}
\includegraphics[width=\textwidth]{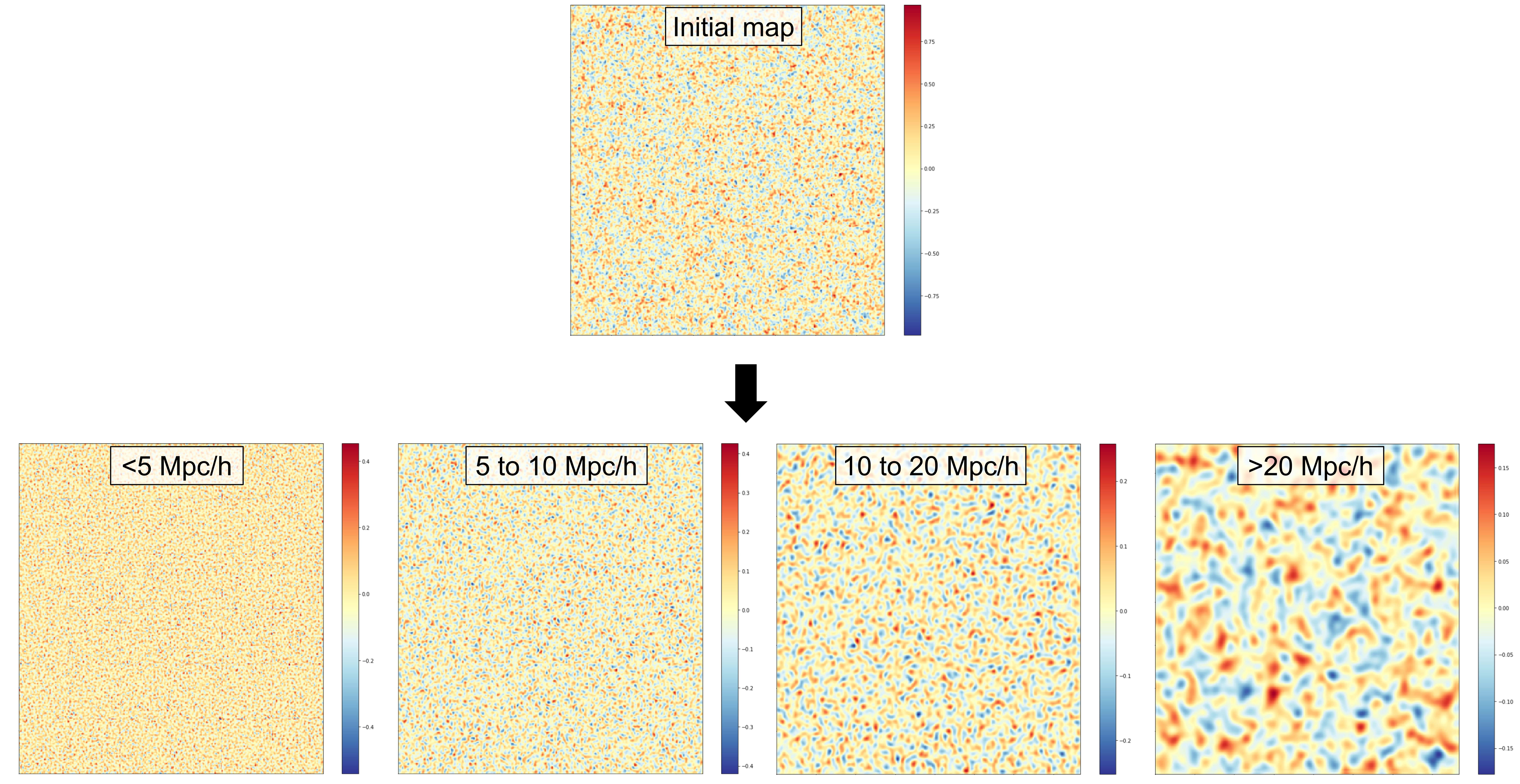}
\caption{Example of a map separated into 4 channels by a starlet transform. Only the first redshift bin is shown, but all four were included in the training. Top row is the initial map, bottom row are the 4 starlet transform channels and the corresponding scales.}
\label{fig:starlet}
\end{figure*}
\begin{figure*}
\includegraphics[width=\textwidth]{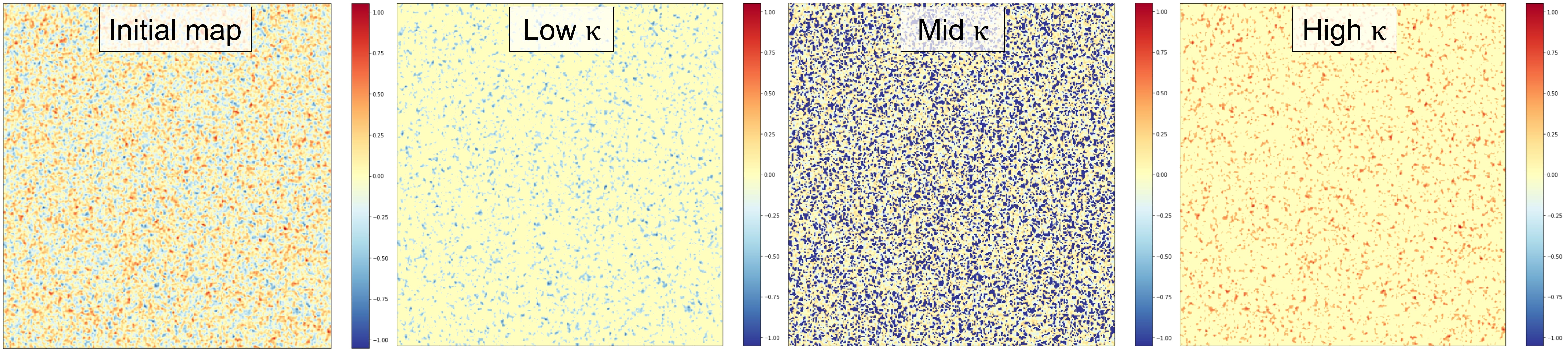}
\caption{Example of a map separated into three convergence regions. Only the first redshift bin is shown, but all four were included in the training.\\ From left to right, there is the base map, the low convergence regions, the mid convergence regions and finally the high convergence regions. Pixels outside of the range appear in yellow for both high and low $\kappa$ and in deep blue for mid $\kappa$.}
\label{fig:threshold}
\end{figure*}
Because this approach requires re-training the networks for each degradation, it has the significant drawback of requiring a long computation time, which is the reason we used a simplified version of the pipeline presented in \cite{kacprzak_deeplss_2022}, but in return allows for a very intuitive understanding of which parts of the surveys hold the most information for the network.\\
\subsection{Data degradations}
To degrade the data, we insert layers in the neural network before the first convolutional layer that modify the data in a number of different ways. Smoothing scales are the only degradations for which the PS-neural network is used. For all other degradations, only the CNN is used as it is the architecture of interest, the PS-neural network being simply a way of implementing a power spectrum analysis.\\

\paragraph{Smoothing scales}
The first degradation is changing the scale at which the Gaussian smoothing of the maps is applied (see Sect.\ref{maps}). Both the CNN and the PS-neural network are trained after smoothing at each of the following scales:
\begin{equation}
   \label{eq:scales}
   R=[2,\ 4\ \textrm{(reference)},\ 8,\ 16]\ \textrm{Mpc/h}
\end{equation}
\paragraph{Starlet transform}
Starlet transforms decompose a given image into $i+1$ channels containing the details at scale $\sim2^i$ pixels for the first $i$ channels and the remainder of the image for the last channel \citep[see][for a full description of starlets transforms]{starck_undecimated_2007}. 
Summing all of the channels recovers the initial image. By performing starlet transforms of the maps, we could present the network with the full information decomposed into multiple channels or select only certain scales.
\paragraph{Convergence thresholds}
Using thresholded Relu layers, we separate the maps into three low, mid and high convergence regions. We chose the threshold to be \kg $=0.25$ which corresponds to a signal to noise ratio of $\sim 5$. In average (this depends on cosmology and redshift) there are roughly 20\% of the total 
pixels in the each of the low or high regions and 60\% of the total pixels in the mid convergence regions. Fig.\ref{fig:threshold} shows an example of a map separated into these 3 convergence regions. Pixels with \kg\ values outside of the chosen range are set to a fixed value, far outside out the range. Different combinations of these regions, presented in the next section, are fed as input to the CNN.
\paragraph{Fourier transform}
Calculating the Fast Fourier Transform (FFT) of the maps, we provide the network with either the full information, but presented in Fourier space, in the form of either real and imaginary parts or amplitude and phase, or just the phase or amplitude information.
\paragraph{Redshift shuffle and sum}
We shuffle the data along the redshift axis, i.e. presenting simulated surveys with a random redshift order, and also sum all four redshift bins into a single map, removing all tomographic information from the survey.
\paragraph{Full pixel shuffling}
Full pixel shuffling allows us to remove all spatial information, essentially being equivalent to an histogram, while still keeping the network architecture identical.
\subsection{Quantifying the network's performance}
\label{performance}
The degradation layers are active during both training and prediction, meaning that the network never has access to the information that is removed. Therefore, if the constraints obtained were still comparable to the one obtained with the reference network (without degradation layers), we can assume that the network does not rely heavily on the removed information or at least that it is redundant and can be replaced in the prediction process.\\
To measure the networks' performance, we focus on two indicators: the standard deviation of the constraint on $S_8=\sigma_8 \sqrt{\Omega_m / 0.3}$, hereafter denoted $\sigma_{S_{8}}$, as well as the information entropy of the MCMC chains. Information entropy is defined in \cite{shannon_mathematical_1948} as:$H= \mathbb{E} [-\log p(X)]$. To calculate the entropy of a MCMC chain, which contains 128k samples that are $[\Omega_m , \sigma_8 ]$ pairs, we split the 2-D prior space into 10k bins and create a histogram of the 128k samples. The entropy is then calculated as:
\begin{equation}
   \label{eq:entropy}
H_{chain}= \ln \frac{1}{n_{bins}} - \left[ \, \sum_{b \in bins} \frac{f_{b}}{n_{samples}} \ln \frac{f_{b}}{n_{samples}} \right]
\end{equation}

where $n_{bins}$=10k is the number of bins, $n_{samples}$=128k is the number of samples and $f_{b}$ is the number of samples that lead to a prediction in the bin $b$.\\
Taking $\ln \frac{1}{n_{bins}}$ as reference means that $H_{chain}$ represents the loss of randomness compared to the prior. Having these two measurements allows for an objective quantification of information loss with each degradation, and both are defined in a "smaller is better" way for easy comparison.\\
We calculate both $\sigma_{S_{8}}$ and $H_{chain}$ for each MCMC chain, then calculate the mean and standard deviation of the two indicators for each model. Note that the means and standard deviations presented thereafter are obtained strictly on the post-processing and do not include any variability in the network training phase. To include that variability, we would have needed to run multiple training for each degradation, which was impossible for computation time reasons. 
\section{Results}
\label{results}
\begin{figure}
\includegraphics[width=\columnwidth]{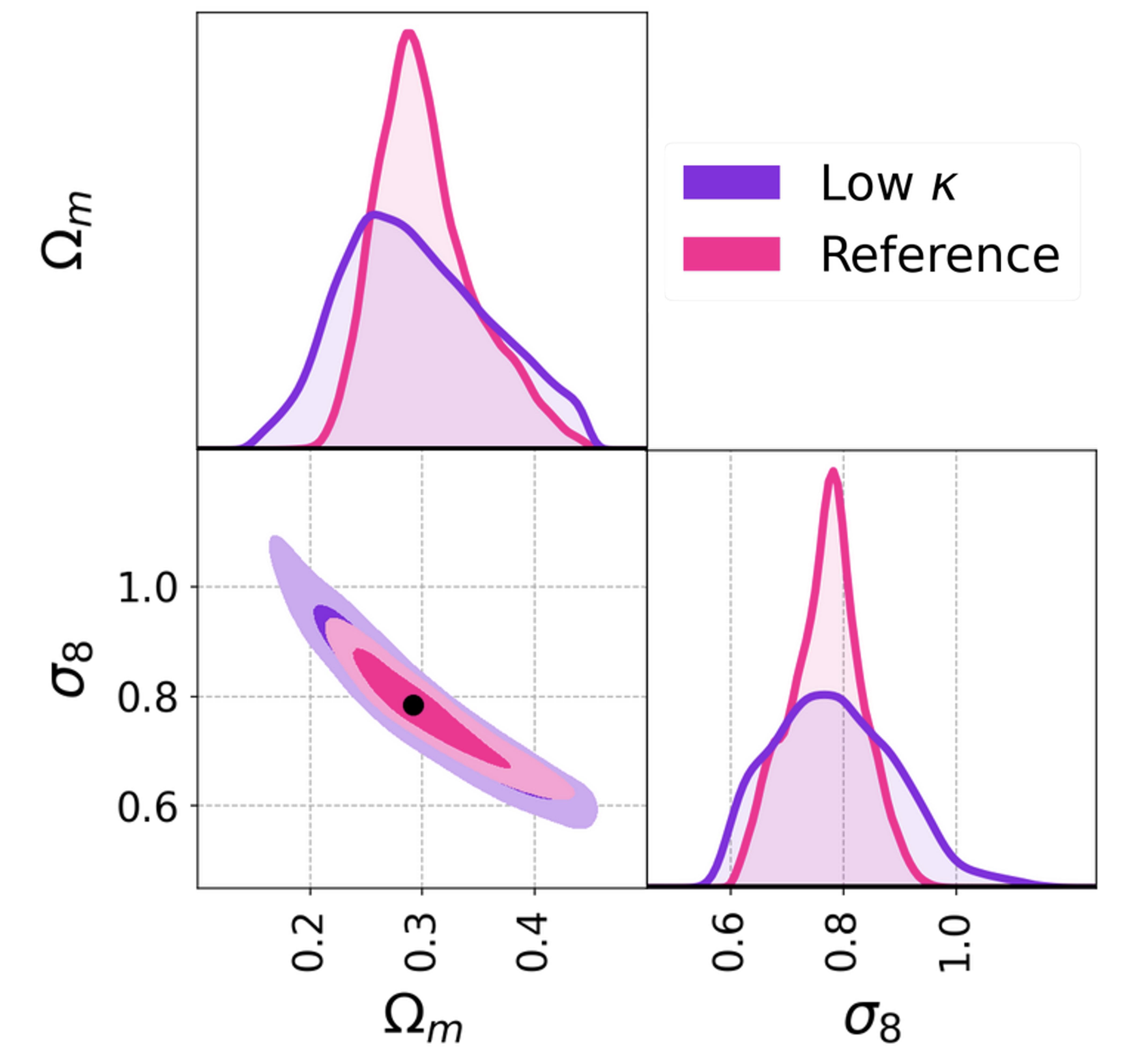}
\caption{Constraints on $[\Omega_m , \sigma_8 ]$ obtained by the CNN. The black dots mark the true value of parameters.}
\label{fig:triangleplot}
\end{figure}

\begin{figure*}
\includegraphics[width=0.96\textwidth]{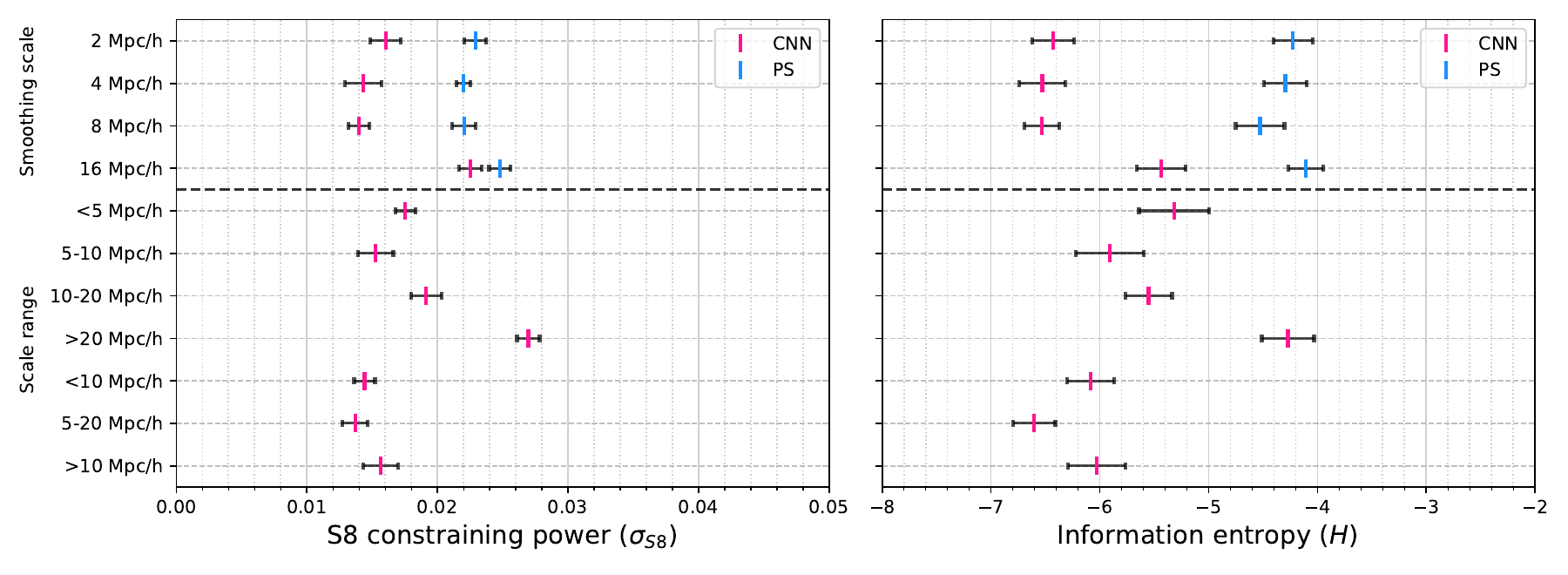}
\caption{Network performance for various scale related degradations. Left panel is $\sigma_{S_{8}}$, the constraining power on $S_8$, right panel is $H$, the information entropy. The top four rows present the performance for various smoothing scales, for both CNN and PS-neural network. The lower rows present the performance of the CNN for various scale range, obtained by keeping only certain starlet transform channels.}
\label{fig:scales}
\end{figure*}

\begin{figure*}
\includegraphics[width=0.96\textwidth]{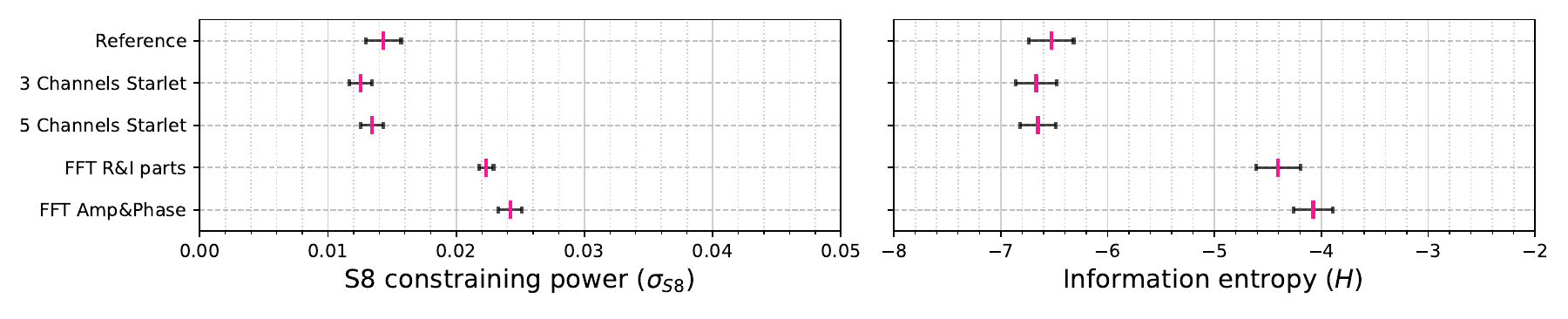}
\caption{CNN performance for various zero-loss transformations. Left panel is $\sigma_{S_{8}}$, the constraining power on $S_8$, right panel is $H$, the information entropy. The results for 3 or 5 channels starlet transform as well as for a Fourier transform in the form of either real and imaginary parts or amplitude and phase are presented.}
\label{fig:nodegrad}
\end{figure*}

\begin{figure*}
\includegraphics[width=0.96\textwidth]{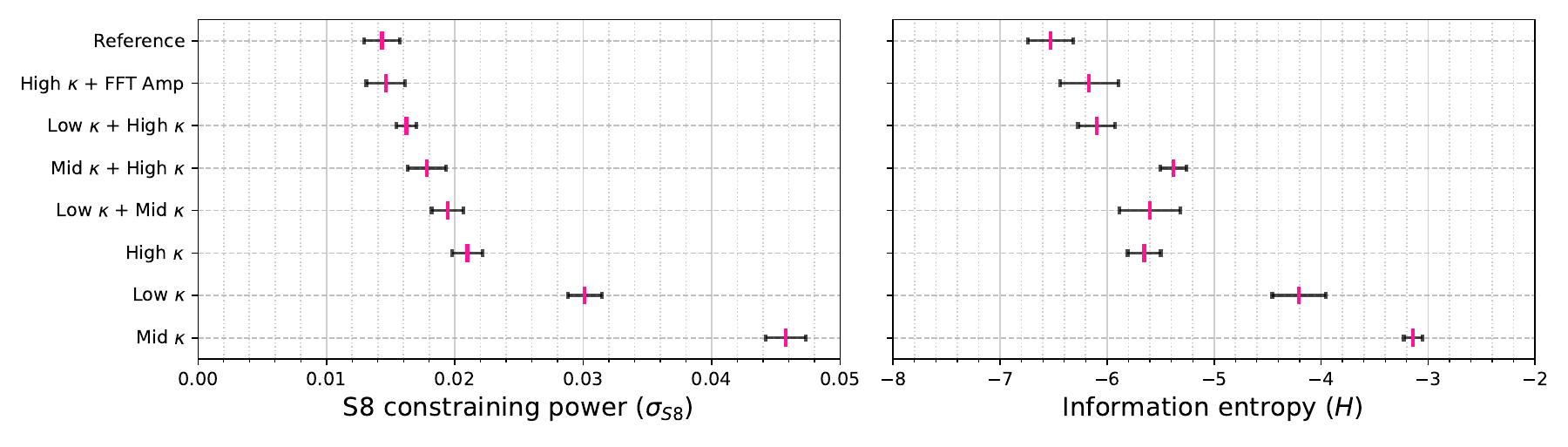}
\caption{CNN performance for various convergence regions selections. Left panel is $\sigma_{S_{8}}$, the constraining power on $S_8$, right panel is $H$, the information entropy. Low/mid/high $\kappa$ denotes the low/mid/high convergence regions. The second row presents the performance of a network taking as input the high convergence regions in one channel and the Fourier transform amplitude in another, to mimic a widely used statistical analysis method: combining the PS and peak counts/Minkowski functionals.}
\label{fig:thresholdperf}
\end{figure*}

\begin{figure*}
\includegraphics[width=0.96\textwidth]{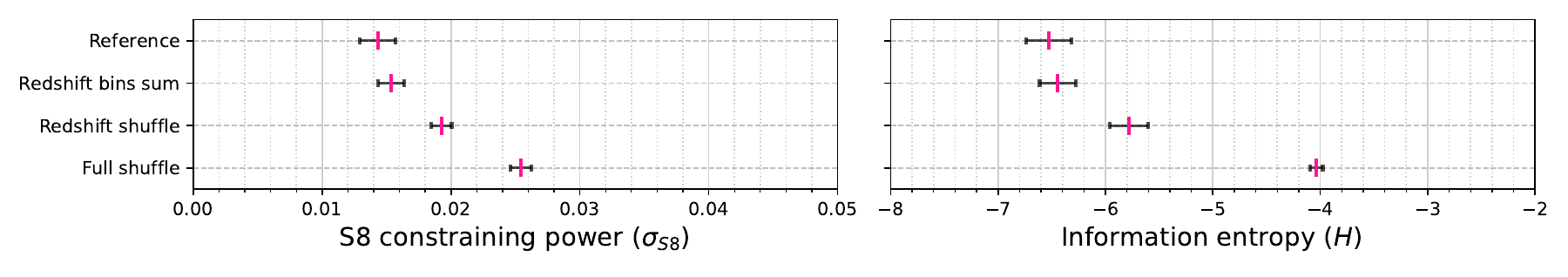}
\caption{CNN performance for redshift shuffling, redshift summing and shuffling all pixels. Left panel is $\sigma_{S_{8}}$, the constraining power on $S_8$, right panel is $H$, the information entropy.}
\label{fig:shuffle}
\end{figure*}

Figures \ref{fig:scales} to \ref{fig:shuffle} present the networks' performance for all the data degradations studied in this work. We note that there is a great coherence between the two indicators: the order of the different degradations in $\sigma_{S_{8}}$ and $H$ is almost identical. The slight disagreements can be explained by one main difference between the two measurements: $\sigma_{S_{8}}$ is not very affected by outliers when compared to the entropy, and probes mostly how tight the centre part of the distribution is.\\
To better visualise the difference in constraint tightness between the different degradations, Fig.\ref{fig:triangleplot} shows a comparison of the constraints obtained with no degradation (Reference, in pink) and with only the low convergence regions (Low $\kappa$, in violet), a degradation that results in some of the worse performance observed in this work. The regions correspond to 68\% and 95\% confidence intervals. 
\subsection{Scales}
Fig.\ref{fig:scales} presents all the results related to scales. We find that there is a very good coherence between the performance when certain scales are removed via smoothing (top four rows) or through removing certain channels of the starlet transforms (bottom rows). The CNN seems very robust to the data being presented in one or multiple channels (this is further explored in Sect.\ref{non_degrading}).\\
From the results, we can deduce that the richest scales in terms of informations are around roughly 8 to 12 Mpc/h: smoothing out smaller scales, likely dominated by noise, doesn't affect the constraining power, and only providing the network with smaller scales results in slightly worse performance. Interestingly, these scales correspond to the transition between linear and non-linear regimes at the considered redshifts, which could explain why they hold so much information.\\
Keeping only very large scales $>$16 Mpc/h results in significantly worse performance than any other scale range with the CNN, but only slightly degrades the PS-neural network's performance: this can be explained by the fact that non-Gaussian information (that the CNN has access to, but the PS does not) is contained in the small, non-linear scales. An interesting feature to note is that removing the smallest ($<$5 Mpc/h) and largest ($>$20 Mpc/h) scales (second-to-last row in Fig.\ref{fig:scales}) seems to slightly improve the CNN's performance although not very significantly. This may be due to the fact that it allows the network to optimize its convolution kernels for the richest scales information wise only and that part of the noise is removed.
\subsection{Non degrading transformations}
\label{non_degrading}
Fig.\ref{fig:nodegrad} presents all the results related to transformations that do not degrade the information, but present it in a different way.\\
Starlet transforms into 3 or 5 channels do not cause any degradation of the CNN's performance, and may even improve it slightly, probably due to the fact that they are a form a data preparation allowing convolution kernels to be specifically optimized to extract information from a certain scale.\\
On the other hand, presenting the data in Fourier space, either as real and imaginary parts or amplitude and phase, as the CNN cannot handle data in the complex space, significantly degrades the networks' ability to extract information. While this is surprising given the fact that no information was removed, this highlights the fact that the CNN architecture is designed to work on real space images.
\subsection{Convergence thresholds}
\label{threshold}
Fig.\ref{fig:thresholdperf} presents all the results obtained when only keeping regions in certain convergence ranges (see Fig.\ref{fig:threshold} for an example of the input maps). We find that the mid convergence regions hold the least information even though they represent a majority of the total area (~60\% of all pixels). This is consistent with what was concluded in \cite{matilla_interpreting_2020} through different methods: pixels with extreme convergence values, i.e. peaks and voids provide significantly more information to the CNN than the regions with intermediate convergence values.\\
Interestingly, even if they provide little information on their own, adding the mid convergence regions back to either low or high convergence regions significantly improves the performance compared to low or high convergence regions alone. Combining the two extreme regions also improves the performance significantly. We can suppose that only one extreme region (especially high convergence regions) provides information mostly on small, non linear scales and lacks information on larger scales: Fig.\ref{fig:threshold} shows how the low and high convergence maps are essentially a collection of small, disconnected peaks and voids with a typical scale of $\sim 7$ Mpc/h. Adding either the mid convergence (where connected regions are much larger) or the opposite extreme likely allows the network to access the information held by larger scales.\\
This hypothesis is also supported by the fact that combining the high convergence regions (i.e. the peaks) with the Fourier transform amplitude, that only contains Gaussian information (like the larger scales), results in essentially indistinguishable performance from the reference model with no data degradation. That result also confirms that the summary statistics approach of combining the power spectrum with peak count and/or Minkowski functionals (morphological descriptors containing information about the peaks/non-Gaussian features) suggested by some studies \citep[e.g.][]{dietrich_cosmology_2010} is likely to result in little information loss.
\subsection{Redshift and spatial information}
Fig.\ref{fig:shuffle} presents all the results obtained when shuffling or summing along the redshift axis or shuffling all the pixel, removing spatial information altogether. Surprisingly, summing the redshift bins results in no significant performance degradation. This is likely due to the simple nature of this work: including astrophysical sources of uncertainty like galaxy intrinsic alignment changes this result, as detailed in Appendix \ref{intrinsic}. Interestingly, shuffling the redshift bins results in worse performance than summing them, even though there is theoretically at least as much information, showing that ill-presented data can lead to prediction errors by the network.\\
Another surprising result is how removing all spatial information does not prevent the network from providing somewhat accurate predictions. The constraining power is comparable to that of the PS-neural network, or to that of the CNN with access to the large scales only.
\section{Discussion and conclusion}
\label{conclusion}
We build on a novel approach to weak lensing surveys analysis through deep learning, via the use of convolutional neural networks, and investigate its behaviour to better understand the algorithm's decision process. We start with a simplified version of the DeepLSS pipeline \citep{kacprzak_deeplss_2022} and add pre-processing layers to the network to remove parts of the information contained in the data set, and measure the constraining performance of the network for various degradations. By analysing how the performance varies, we try to understand what parts and features of the surveys are most important in the network's prediction process.\\
Using the approach from \cite{fluri_cosmological_2019}, \cite{fluri_full_2022}, and \cite{kacprzak_deeplss_2022}, we run N-body simulations of the Universe for various cosmological parameters values and use them to generate tomographic weak lensing convergence \kg\ maps via a pencil beam approach. We create two networks to analyse the data: a residual convolutional neural network that takes the raw maps as input and a power spectrum neural network that first calculates all auto- and cross- spectra of the maps then passes them through a simple neural network. Both architectures are trained using likelihood loss between outputs and true input parameters to compress the data into our two cosmological parameters of interest, \omegm\ and \sig8, and their covariances. We then obtain constraints via Bayesian analysis with a Markov Chain Monte Carlo sampler.\\
To understand the networks' prediction process, we measure its performance with two indicators: $\sigma_{S_8}$, the tightness of constraint on $S_8$, and $H$, the information entropy of the posterior. We first set a reference by running the convolutional neural network on the unmodified data set. We then introduce layers to the network to measure its performance on degraded data with some information removed. We removed certain scales, separated the data into low/mid/high convergence regions, and removed redshift or spatial information.\\
We find that the scales that seems to hold the most information are 8 to 12 Mpc/h, which corresponds to the transition between linear and non-linear scales. As long as these scales are present in the data, performance is indistinguishable from the reference. Large scales only (over 16 Mpc/h) lead to significantly worse performance, similar to the power spectrum approach. Smaller scales only (below 10 Mpc/h) lead to slightly worse performance than reference.\\
We report that separating the data into multiple channels via starlet transform does not affect performance, but presenting the data in Fourier space significantly affects the constraining power.\\
We found that mid convergence regions hold little information on their own compared to low or high convergence regions (i.e. voids or peaks), but combining mid with either low or high convergence regions significantly improved the performance compared to either extreme on its own. This is likely due to the fact that low or high convergence maps lack large connected regions, and contain mainly non-Gaussian information. We also report that combining high convergence regions with the Fourier transform amplitude leads to the same performance as the reference run, confirming that combining power spectrum and information on the peaks \citep[as suggested by e.g.][]{dietrich_cosmology_2010} is a very good approach, and supporting our hypothesis that the lack of Gaussian information is holding back the network's performance when provided with either only peaks or voids.\\
Summing the redshift bins to create a non-tomographic survey surprisingly does not affect the performance much. We attribute this behaviour to our simplified baseline physical modelling that lacks astrophysical uncertainties like intrinsic alignment, as including intrinsic alignment changes this results (see Appendix \ref{intrinsic}). Removing the spatial information altogether by shuffling the pixels leads to a significant performance loss, but the network is still able to make relevant predictions, with performance similar to a power spectrum approach.\\
Direct comparison with other studies is difficult due to the novelty of the field of interpretability of deep learning methods applied to weak-lensing surveys and the diversity of methods used, but we can compare our results to the work of \cite{matilla_interpreting_2020}, who also tried to interpret the decision process of a CNN trained on weak lensing data, albeit with a very different method. Using saliency maps to analyse the decision process of an already trained network, they found that the CNN relied heavily on the pixels with the most extreme values in the convergence maps, which is consistent with the results we present in Sect.\ref{threshold}. When studying noiseless maps, they found that pixels with the lowest convergence were most relevant to the decision, which doesn't agree with our findings. Nevertheless, when adding shape noise, like what is done in this work, they found that the highest convergence regions were the most important in the network's decision process, which aligns with our results. Saliency maps were also studied with the DeepLSS architecture in the original study \citep{kacprzak_deeplss_2022}, but a direct comparison with this work is not straightforward as both convergence and galaxy density maps were used as combined probes in the original study, and the sensitivity study mainly indicated a correlation between the importance of pixels in one probe and their value in the other probe.\\
The key conclusion from this work is that, much like fixed summary statistics analysis methods, the convolutional neural network's constraining performance is strongest when provided with both Gaussian and non-Gaussian information, and that structures at the limit between linear and non-linear regimes are particularly important in its prediction process.

%
%

\bibliography{fullbib}
\bibliographystyle{aa}

\begin{appendix}
\section{Results including intrinsic alignment}
\label{intrinsic}
In this appendix, we present the results obtained when modelling galaxy intrinsic alignment in the convergence maps. Intrinsic alignment is a source of systematic error in weak lensing surveys that arises from the fact that galaxies are not randomly oriented in the Universe, but could be aligned with each other due to large-scale tidal fields. This leads to a correlation between the shapes of galaxies and the weak lensing signal, which can be a significant source of bias in the analysis of weak lensing surveys \citep{hirata_intrinsic_2004}. Additionally, the magnitude of this effect depends on the physics of galaxy formation, and is very difficult to estimate from either simulations or theoretical calculations. Given the fact that the aim of this work is not provide constraints or even forecast constraining power, we chose to present the main analysis without including this effect, and verify here whether any conclusions are changed with its inclusion.\\
To model intrinsic alignment we follow the method used in \cite{kacprzak_deeplss_2022} \citep[originally derived from][]{zurcher_dark_2022}, based on a non-linear alignment model \citep{bridle_dark_2007,joachimi_constraints_2011}. We create intrinsic alignment $\kappa_{IA}$ maps for each redshift bin following Eq.\ref{eq:born_approx} in a similar way to the convergence maps, except that the kernel is the intrinsic alignment kernel, given by:
\begin{equation}
   \label{eq:ia_weights}
   W^{IA}_{b} = \frac{\int_{\Delta z_b}^{} dz F(z) n(z)} {\int_{\Delta z_b}^{} \frac{dz}{E(z)} \int_{z_0}^{z_s} dz' n(z')},
\end{equation}
where $n(z)$ is the redshift distribution of galaxies in a given redshift bin, $z_s$ and $z_0$ are the source and observer redshifts, and $F(z)$ is a cosmology dependent term:
\begin{equation}
   \label{eq:ia_F}
   F(z) = -C_1 \rho_{\text{crit}} \frac{\Omega_m}{D_{+}(z)},
\end{equation}
where $C_1=5.10^{-14} h^{-2} M_{\odot} \, \text{Mpc}^3$, $\rho_{\text{crit}}$ is the critical density of the Universe, and $D_{+}(z)$ is the linear growth factor (normalized to $D_{+}(0)=1$). Finally, the mean of each intrinsic alignment maps is removed.\\
The intrinsic alignment maps are then added to the convergence maps on the fly during the training and prediction process, using a single effective scaling value per redshift bin $i$:
\begin{equation}
   \label{eq:ia_sum}
   \kappa_{g}^i = \kappa_{g}^i + A^i_{IA} \kappa_{IA}^{(i)},
\end{equation}
with
\begin{equation}
   \label{eq:ia_scaling}
   A^i_{IA} = A_{IA}\int_{z}^{} dz n^i(z)\left(\frac{1+z}{1+z_0}\right)^{\eta_{A_{IA}}},
\end{equation}
where $n^i(z)$ is the redshift distribution of galaxies in the redshift bin $i$, $A_{IA}$ is the intrinsic alignment amplitude and $\eta_{A_{IA}}$ is the intrinsic alignment redshift evolution. When intrinsic alignment is included, $A_{IA}$ and $\eta_{A_{IA}}$ are treated as parameters of the global model, and are varied during training and predicted by the network along with \omegm and \sig8 during the prediction step. Large flat priors are used for both parameters ($[-6,6]$ for $A_{IA}$ and $[-4,6]$ for $\eta_{A_{IA}}$).\\
 \begin{figure*}[!ht]
   \includegraphics[width=0.96\textwidth]{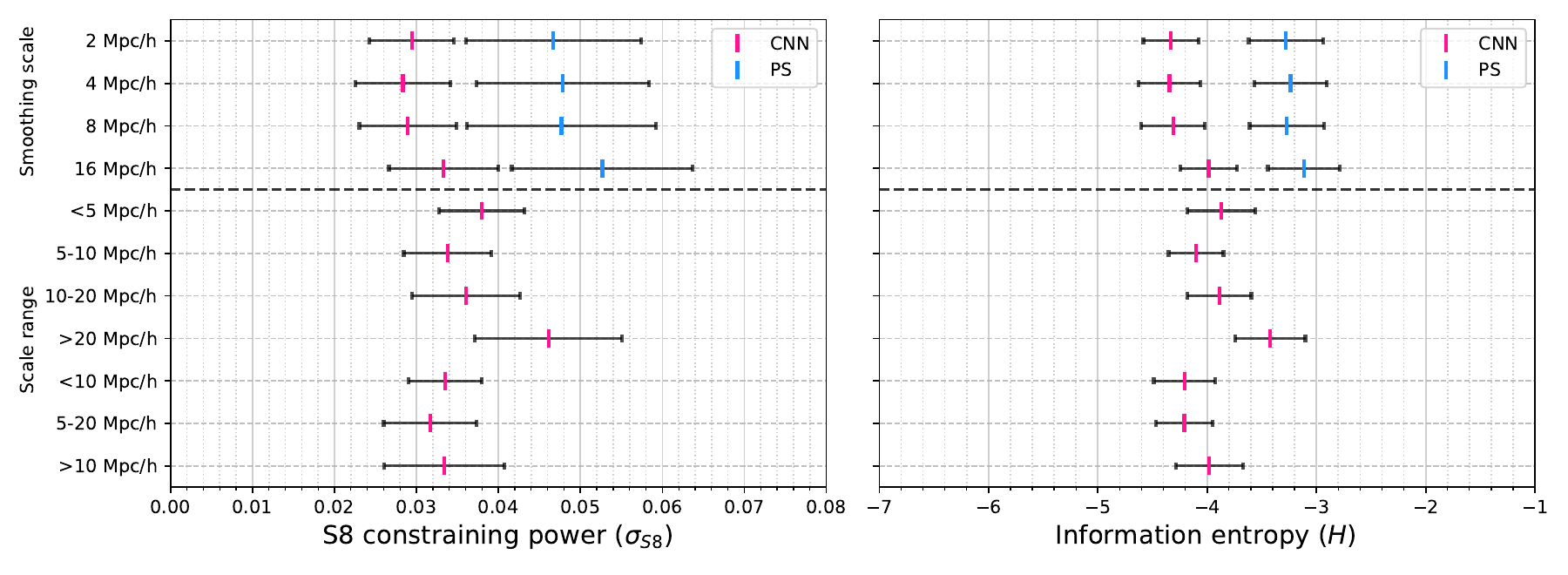}
   \caption{Network performance for various scale related degradations. Left panel is $\sigma_{S_{8}}$, the constraining power on $S_8$, right panel is $H$, the information entropy. The top four rows present the performance for various smoothing scales, for both CNN and PS-neural network. The lower rows present the performance of the CNN for various scale range, obtained by keeping only certain starlet transform channels.}
   \label{fig:scalesAIA}
   \end{figure*}
\begin{figure*}[!ht]
   \includegraphics[width=0.96\textwidth]{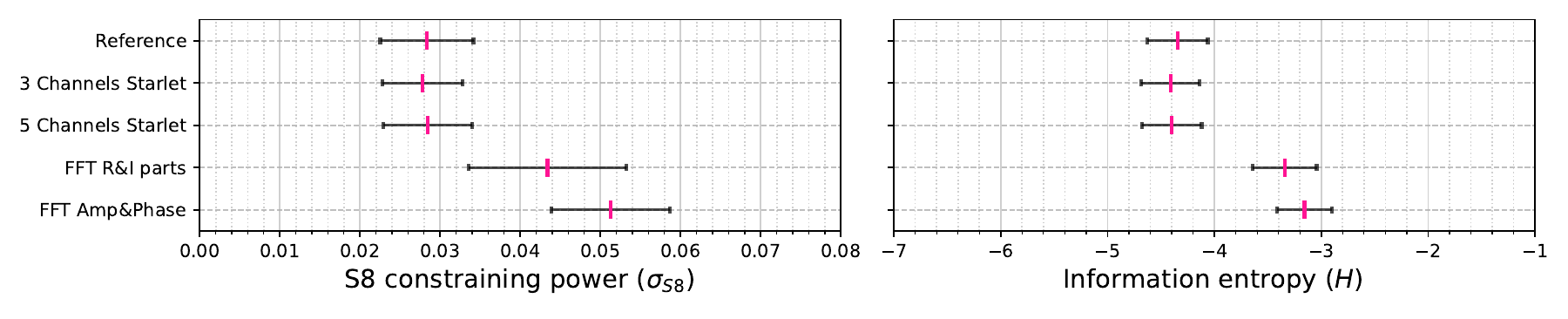}
   \caption{CNN performance for various zero-loss transformations. Left panel is $\sigma_{S_{8}}$, the constraining power on $S_8$, right panel is $H$, the information entropy. The results for 3 or 5 channels starlet transform as well as for a Fourier transform in the form of either real and imaginary parts or amplitude and phase are presented.}
   \label{fig:nodegradAIA}
   \end{figure*}
\begin{figure*}[!ht]
   \includegraphics[width=0.96\textwidth]{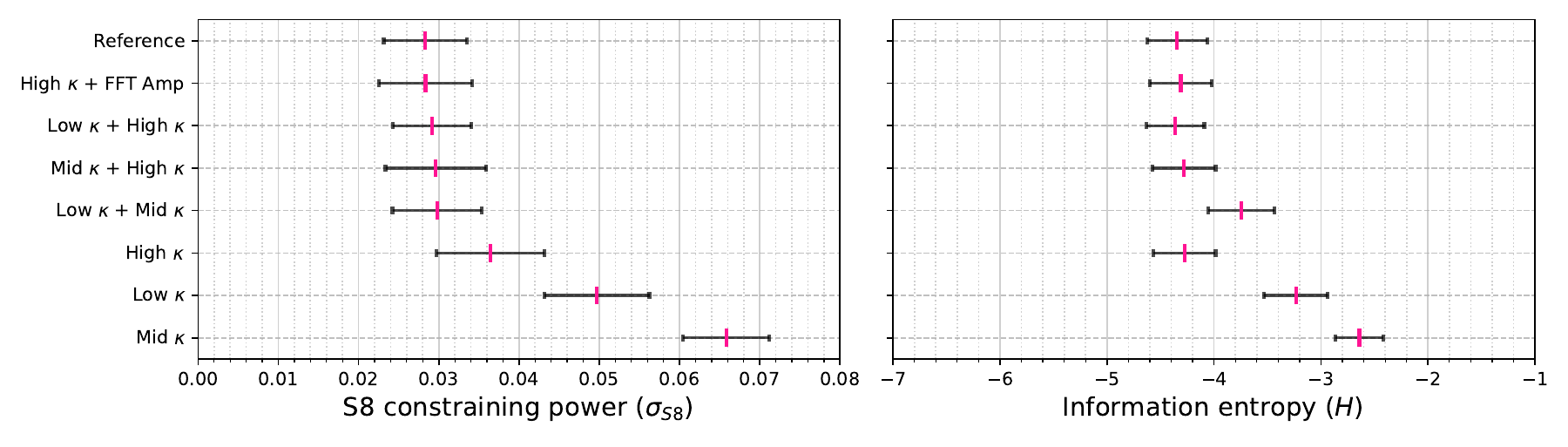}
   \caption{CNN performance for various convergence regions selections. Left panel is $\sigma_{S_{8}}$, the constraining power on $S_8$, right panel is $H$, the information entropy. Low/mid/high $\kappa$ denotes the low/mid/high convergence regions. The second row presents the performance of a network taking as input the high convergence regions in one channel and the Fourier transform amplitude in another, to mimic a widely used statistical analysis method: combining the PS and peak counts/Minkowski functionals.}
   \label{fig:thresholdperfAIA}
   \end{figure*}
\begin{figure*}[!ht]
   \includegraphics[width=0.96\textwidth]{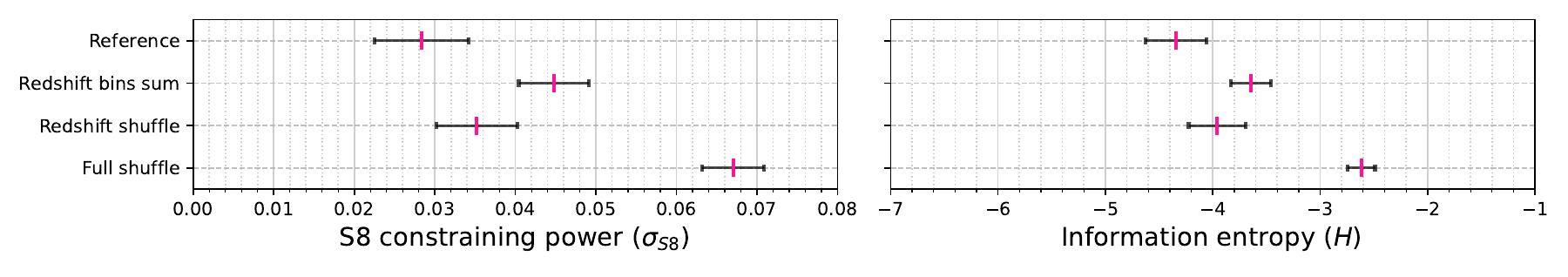}
   \caption{CNN performance for redshift shuffling, redshift summing and shuffling all pixels. Left panel is $\sigma_{S_{8}}$, the constraining power on $S_8$, right panel is $H$, the information entropy.}
   \label{fig:shuffleAIA}
   \end{figure*}
   Fig.\ref{fig:scalesAIA} to \ref{fig:shuffleAIA} present the networks' performance for all the data degradations studied in this work, but with the addition of intrinsic alignment. We find that the results are in general agreement to those obtained without intrinsic alignment, with overall worse constraining power and less differences between the different degradations. The small scales seem less important in this case, with the constraining power not diminishing as much as in the baseline analysis when the smoothing scale is increased. The other significant difference is that the performance of the network when summing the redshift bins is now significantly worse than the reference model. This is likely due to the fact that the network is unable to correctly fit for intrinsic alignment without redshift information, as the evolution of intrinsic alignment with redshift is one of its potential distinguishing features with cosmic shear.\\
\end{appendix}
\end{document}